\documentclass[amsmath,amssymb,aps,prl,11pt,tightenlines,superscriptaddress,
nofootinbib,preprintnumbers,notitlepage]{revtex4-1}

\usepackage[T1]{fontenc}
\usepackage[utf8]{inputenc}
\usepackage[british]{babel}

\usepackage{color,xcolor}

\definecolor{dark-red}{rgb}{0.8, 0.0, 0.1803921568627451}
\definecolor{dark-blue}{rgb}{0.0, 0.0, 0.803921568627451}
\definecolor{dark-green}{rgb}{0.0, 0.39215686274509803, 0.0}
\definecolor{dark-orange}{rgb}{0.8, 0.4, 0.0}

\usepackage{amsmath,dsfont,mathtools}
\usepackage{siunitx}
\usepackage{graphicx}
\usepackage{multirow,booktabs,tabularx,float}
\usepackage[font=normalsize]{subfig}
\usepackage{slashed,braket}
\usepackage{feynmp-auto}
\usepackage[sort&compress]{natbib}
\PassOptionsToPackage{hyphens}{url}\usepackage[pdftex,hidelinks,linkcolor={dark-blue}, citecolor={dark-green}, urlcolor={dark-red},bookmarksopen,linktoc=all]{hyperref}
\usepackage{enumitem} 
\usepackage[stretch=15,shrink=15]{microtype} 
\usepackage{tabto}
\usepackage{accents} 
\usepackage{xspace} 

\newcommand{\toolfont}[1]{\textsc{#1}}

\newcommand{\im}{\mathrm{i}}
\newcommand{\diff}{\mathrm{d}}

\newcommand{\ie}{{i.\,e.}~}

\newcommand{\lgr}[1]{\mathcal{L}_\text{#1}}

\newcommand{\phisq}{\phi^\dagger \phi}

\newcommand{\sally}{\textsc{Sally}\xspace}
\newcommand{\sallino}{\textsc{Sallino}\xspace}

\newcommand{\rascal}{\textsc{Rascal}\xspace}

\newcommand{\intractablep}{\ensuremath {\textcolor{dark-red}{p}}}
\newcommand{\intractabler}{\ensuremath {\textcolor{dark-red}{r}}}

\newcommand{\intractablet}{\ensuremath {\textcolor{dark-red}{t}}}
\newcommand{\intractablez}{\ensuremath {\textcolor{dark-red}{Z}}}
\newcommand{\localmodel}{\ensuremath {\textcolor{dark-red}{p_{\text{local}}}}}

\newcommand{\intz}{\int \! \diff z\;}

\newlength{\hhatheight}

\DeclareMathOperator*{\argmin}{arg\,min}

\newcolumntype{R}{>{\raggedleft\arraybackslash}X}%
\newcolumntype{L}{>{\raggedright\arraybackslash}X}%

\AtBeginDocument{
  \heavyrulewidth=.08em
  \lightrulewidth=.05em
  \cmidrulewidth=.03em
  \belowrulesep=.65ex
  \belowbottomsep=0pt
  \aboverulesep=.4ex
  \abovetopsep=0pt
  \cmidrulesep=\doublerulesep
  \cmidrulekern=.5em
  \defaultaddspace= .5em
  \setlength{\tabcolsep}{0.5em}
}

\setlist[itemize]{itemsep=1pt,parsep=1pt, topsep=1pt}

\begin{document}


\title{Constraining Effective Field Theories with Machine Learning}

\author{Johann Brehmer}
\affiliation{New York University, USA}

\author{Kyle Cranmer}
\affiliation{New York University, USA}

\author{Gilles Louppe}
\affiliation{University of Li\`{e}ge, Belgium}

\author{Juan Pavez}
\affiliation{Federico Santa Mar\'ia Technical University,  Chile}

\date{\today}

\begin{abstract}
  We present powerful new analysis techniques to constrain effective field theories at the LHC. By leveraging the structure of particle physics processes, we extract extra information from Monte-Carlo simulations, which can be used to train neural network models that estimate the likelihood ratio. These methods scale well to processes with many observables and theory parameters, do not require any approximations of the parton shower or detector response, and can be evaluated in microseconds. We show that they allow us to put significantly stronger bounds on dimension-six operators than existing methods, demonstrating their potential to improve the precision of the LHC legacy constraints.
\end{abstract}

\maketitle

\section{Introduction}
\label{sec:intro}

Precision constraints on indirect signatures of physics beyond the Standard Model (SM) will be an important part of the legacy of the Large Hadron Collider (LHC) experiments. A key component of this program are limits on the dimension-six operators of the SM Effective Field Theory (SMEFT)~\cite{Buchmuller:1985jz, Grzadkowski:2010es}. Processes relevant to these measurements are often sensitive to a large number of EFT coefficients, which predict subtle kinematic signatures in high-dimensional phase spaces.

Traditionally, such signatures are analysed by focussing on a few hand-picked kinematic variables. This approach discards any information in the remaining directions of phase space. Well-chosen variables typically yield precise bounds along individual directions of the parameter space, but only weak constraints in other directions~\cite{Brehmer:2016nyr, Brehmer:2017lrt}. The sensitivity to multiple parameters can be substantially improved by using the fully differential cross section. This is the forte of the Matrix Element Method~\cite{Kondo:1988yd, Abazov:2004cs, Artoisenet:2008zz, Gao:2010qx, Alwall:2010cq, Soper:2011cr, Soper:2012pb, Bolognesi:2012mm, Avery:2012um, Andersen:2012kn, Campbell:2013hz, Artoisenet:2013vfa, Gainer:2013iya, Soper:2014rya, Schouten:2014yza, Englert:2015dlp, Martini:2015fsa, Gritsan:2016hjl, Martini:2017ydu} and Optimal Observables~\cite{Atwood:1991ka, Davier:1992nw, Diehl:1993br} techniques, which are based on the parton-level structure of a given process. But these methods either neglect or approximate the parton shower and detector response. Moreover, even a simplified description of the detector effects requires the numerically expensive evaluation of complicated integrals for each observed event. None of these established approaches scales well to high-dimensional problems with many parameters and observables, such as the SMEFT measurements.

Recently, we have developed new techniques to constrain continuous theory parameters in LHC experiments based on machine learning and neural networks. The companion publication~\cite{companion_long} is an extensive guide that thoroughly describes and compares a number of different techniques for this problem. In addition, Ref.~\cite{companion_nips} presents the methods in a more abstract setting. Here we want to highlight the key idea: by harnessing the structure of particle physics processes, we can extract additional information from Monte-Carlo simulations that characterizes the dependence of the likelihood on the theory parameters. This augmented data can be used to train neural networks that precisely estimate likelihood ratios, the preferred test statistics for limit setting at the LHC. We sketch two particularly useful algorithms based on these ideas and demonstrate their performance in the example process of weak-boson-fusion Higgs production in the four-lepton decay mode.

\section{Techniques}
\label{sec:methods}

\subsection{Learning likelihood ratios}
\label{sec:rascal}

Constraints on beyond-the-standard-model theories by the LHC experiments are typically based on likelihood ratio tests, as they enjoy many optimal statistical properties. In particle physics processes, the likelihood $\intractablep(x | \theta)$ of theory parameters $\theta$ given data $x$ typically factorizes into a parton-level process, which depends on the theory parameters, followed by the parton shower and detector interactions:
\begin{equation}
  \intractablep(x | \theta)
  = \int  \! \diff z_{\text{detector}} \, \int \! \diff z_{\text{shower}} \, \intz
  \underbrace{
    \intractablep(x | z_{\text{detector}}) \; \intractablep(z_{\text{detector}} | z_{\text{shower}}) \;
    \intractablep(z_{\text{shower}} | z) \; p (z | \theta)
  }_{= \intractablep(x, z_{\text{detector}}, z_{\text{shower}}, z | \theta)} \,.
  \label{eq:latent_structure}
\end{equation}
Here $p(z|\theta) = 1/\sigma(\theta) \, \diff \sigma (\theta) / \diff z$ is the probability density of the parton-level momenta $z$ conditional on the theory parameters $\theta$. The other conditional densities $\intractablep(z_{\text{shower}} | z)$,  $\intractablep(z_{\text{detector}} | z_{\text{shower}})$, and $\intractablep(x | z_{\text{detector}})$ describe how parton-level four-momenta $z$ evolve to reconstruction-level observables $x$ through the parton shower, detector effects, and the reconstruction procedure.

Simulators such as \toolfont{Pythia}~\cite{Sjostrand:2007gs} and \toolfont{Geant4}~\cite{Agostinelli:2002hh} use Monte-Carlo techniques to sample from these distributions. Each step of this chain only depends on the previous one. Our analysis techniques will rely on this Markov property. The simulation of a single event can easily involve many millions of random variables, it is infeasible to explicitly calculate the integral over this enormous space. This is why the likelihood function and the likelihood ratio are intractable, \ie they cannot be evaluated for a given $x$ and $\theta$. We denote this intractability with red symbols. An optimal analysis strategy thus requires a precise estimator of the likelihood ratio based on the available data from the simulator.

Crucially, though, evaluating the density $p(z|\theta)$ of parton-level four-momenta is tractable: the matrix element and the parton density functions can be evaluated for arbitrary four-momenta $z$ and parameter values $\theta$. Matrix-element codes define functions that return the squared matrix element for a given phase-space point $z$.

This property allows us to extract more information from the simulator than just the generated samples of observables $\{x\}$: we can access the corresponding parton-level momenta $\{z\}$ and extract the \emph{joint likelihood ratio}
\begin{align}
  r(x, z | \theta_0 , \theta_1) &\equiv \frac {\intractablep(x, z_{\text{detector}}, z_{\text{shower}}, z | \theta_0)} {\intractablep(x, z_{\text{detector}}, z_{\text{shower}}, z | \theta_1)} \notag \\
&= \frac {\intractablep(x | z_{\text{detector}})} {\intractablep(x | z_{\text{detector}})} \, \frac {\intractablep(z_{\text{detector}} | z_{\text{shower}})}  {\intractablep(z_{\text{detector}} | z_{\text{shower}})} \, \frac {\intractablep(z_{\text{shower}} | z)}  {\intractablep(z_{\text{shower}} | z)} \, \frac {p(z | \theta_0)} {p(z | \theta_1)}
= \frac {p(z | \theta_0)} {p(z | \theta_1)} \,,
\end{align}
as well as the \emph{joint score}
\begin{equation}
  t(x, z | \theta_0)
  \equiv \nabla_\theta \log \intractablep(x, z_{\text{detector}}, z_{\text{shower}}, z | \theta)  \Biggr |_{\theta_0}
  = \frac {\nabla_\theta p(z | \theta)} {p(z |  \theta)} \Biggr |_{\theta_0} \,,
\end{equation}
which describes the relative gradient of the likelihood with respect to theory parameters. Because all intractable parts of the likelihood cancel in the ratio, this step does not require any assumptions or approximations about shower and detector.

These joint quantities $r(x, z | \theta_0, \theta_1)$ and $t(x, z | \theta_0)$ depend on the parton-level momenta $z$, which are of course not available for measured data. Their connection to the likelihood ratio $\intractabler(x | \theta_0, \theta_1)$ that we are interested in is not obvious (essentially because the integral of the ratio is not the ratio of two integrals). However, in Ref.~\cite{companion_nips} we show that they can be used to define functionals $L_r[g]$ and $L_t[g]$ that are extremized by the likelihood ratio
\begin{equation}
  \intractabler(x| \theta_0, \theta_1) \equiv \frac {\intractablep(x | \theta_0)} {\intractablep(x | \theta_1)}
  = \argmin_{g} L_r[g]
\end{equation}
and the score
\begin{equation}
  \intractablet(x | \theta_0) \equiv \nabla_\theta \log \intractablep(x | \theta) \Biggr |_{\theta_0}
  = \argmin_g L_t[g] \,,
  \label{eq:score}
\end{equation}
respectively.

\begin{figure}
  \centering%
  \includegraphics[width=\textwidth]{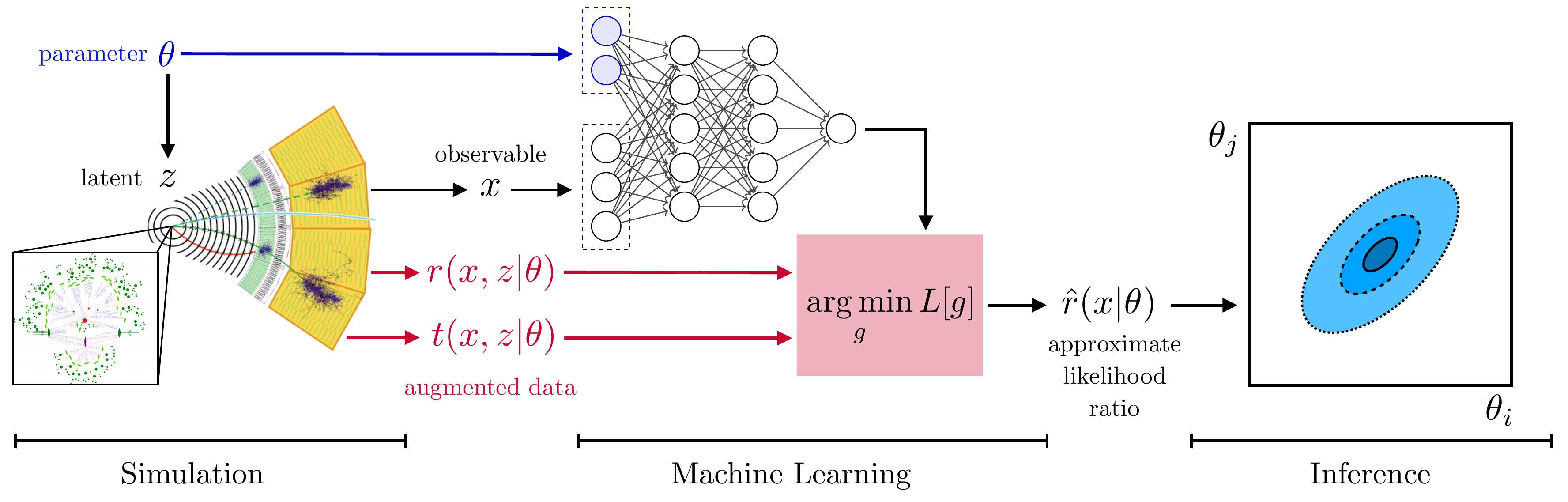}%
  \caption{Schematic overview of the techniques presented in this Letter.$^2$}
  \label{fig:schematic}
\end{figure}

We implement this approach through machine learning, approximating the functionals $L_r[g]$ and $L_t[g]$ through suitable loss functions based on data available from the simulator, see Fig.~\ref{fig:schematic}. The extremization of the loss functional is estimated by training a deep neural network using stochastic gradient descent on the network's parameters.

Based on this idea, we define the \rascal\footnote{\textbf{R}atio \textbf{a}nd \textbf{sc}ore \textbf{a}pproximate \textbf{l}ikelihood ratio} technique that uses both pieces of information\,--\,the joint likelihood ratio and the joint score\,--\,simultaneously to train an estimator $\hat{r}(x | \theta_0, \theta_1)$ for the likelihood ratio. This approach is essentially a machine-learning version of the Matrix Element Method. It replaces computationally expensive numerical integrals with an upfront regression phase, after which the likelihood ratio can be evaluated in microseconds per event and parameter point. Instead of manually specifying simplified smearing functions, the effect of parton shower and detector is learned from full simulations. By using all available information from the simulator, this estimator maximizes the fidelity of the likelihood ratio estimation (and therefore the precision of measurements), at the cost of a somewhat complex architecture. 
\footnotetext{Parts of the figure are based on Ref.~\cite{Barney:2120661} and on an image created by Frank Krauss.}

\subsection{Local approximation}
\label{sec:sally}

In the neighborhood of the Standard Model (or any other reference point), we can approximate the score $\intractablet(x | \theta)$ as independent of $\theta$, and Eq.~\eqref{eq:score} is solved by
\begin{equation}
  \localmodel(x|\theta) = \frac 1 {\intractablez(\theta)} \, \intractablep(t(x|\theta_{SM})\,|\,\theta_{SM}) \, \exp[ \intractablet(x|\theta_{SM}) \cdot (\theta - \theta_{SM}) ]
  \label{eq:local_model}
\end{equation}
with a normalisation factor $\intractablez(\theta)$.

This local model is in the exponential family of probability distributions. The score $\intractablet(x| \theta_{SM})$ are the sufficient statistics, \ie functions of the observables that contain all the information on $\theta$. A precise score estimator $\hat{t}(x | \theta_{SM})$ therefore defines a vector of ideal, loss-free summary statistics,\footnote{Independently, the role of the score was studied for cosmological data~\cite{Alsing:2018eau}.} at least in the proximity of the Standard Model. The estimated score is essentially a machine-learning version of Optimal Observables.

In the companion paper, we construct an estimator for the score based on the availability of the joint score from the simulator discussed above, again realized as a neural network. This is the basis of the new \sally\footnote{\textbf{S}core \textbf{a}pproximates \textbf{l}ikelihood \textbf{l}ocall\textbf{y}} method to estimate likelihood ratios.

In fact, this dimensionality reduction can be taken one step further. The scalar product
\begin{equation}
  \hat{h}(x | \theta_0, \theta_1) \equiv \hat{t}(x | \theta_{SM}) \cdot (\theta_0 - \theta_1)
\end{equation}
 encapsulates all the discrimination power between $\theta_0$ and $\theta_1$, at least in the local model approximation. This allows us to compress high-dimensional observations to a single scalar function without losing any sensitivity, even for hundreds of theory parameters. In Ref.~\cite{companion_long} we define the  \sallino\footnote{\textbf{S}core \textbf{a}pproximates \textbf{l}ikelihood \textbf{l}ocally \textbf{in} \textbf{o}ne direction} technique for likelihood ratio estimation based on this dimensionality reduction.

By construction, the \sally and \sallino techniques work very well close to the Standard Model. While the local model approximation may deteriorate far away from the Standard Model, the effect of this approximation error is reduced sensitivity and weaker bounds\,---\,it does not lead to overly optimistic results. These approaches are simple and robust, and in particular the \sallino method scales exceptionally well to high-dimensional parameter spaces.

\section{Example process}
\label{sec:example}

We demonstrate these two methods by calculating expected SMEFT constraints based on the kinematics of Higgs production in weak boson fusion in the four-lepton mode. This process is particularly sensitive to two operators~\cite{Brehmer:2016nyr, Brehmer:2017lrt}
\begin{equation}
  \lgr{} = \lgr{SM}
  + \frac {f_{W}} {\Lambda^2} \; \dfrac{\im g}{2} \, (D^\mu\phi)^\dagger \, \sigma^a \, D^\nu\phi \; W_{\mu\nu}^a \;
  {} - \frac {f_{WW}} {\Lambda^2} \; \frac{g^2}{4} \, (\phisq) \; W^a_{\mu\nu} \, W^{\mu\nu\, a} \,.
\end{equation}
We generate event samples using a combination of \toolfont{MadGraph~5}~\cite{Alwall:2014hca} and its add-on \toolfont{MadMax}~\cite{Cranmer:2006zs, Plehn:2013paa, Kling:2016lay}. In order to be able to assess the performance of our methods, we use an idealized setup in which the momenta of the partons can be measured exactly, so that we can compare the results to the true likelihood ratio. In Ref.~\cite{companion_long} we describe the setup in more detail and show results for a more realistic simulation.

\begin{figure}
  \centering%
  \includegraphics[width=0.49 \textwidth]{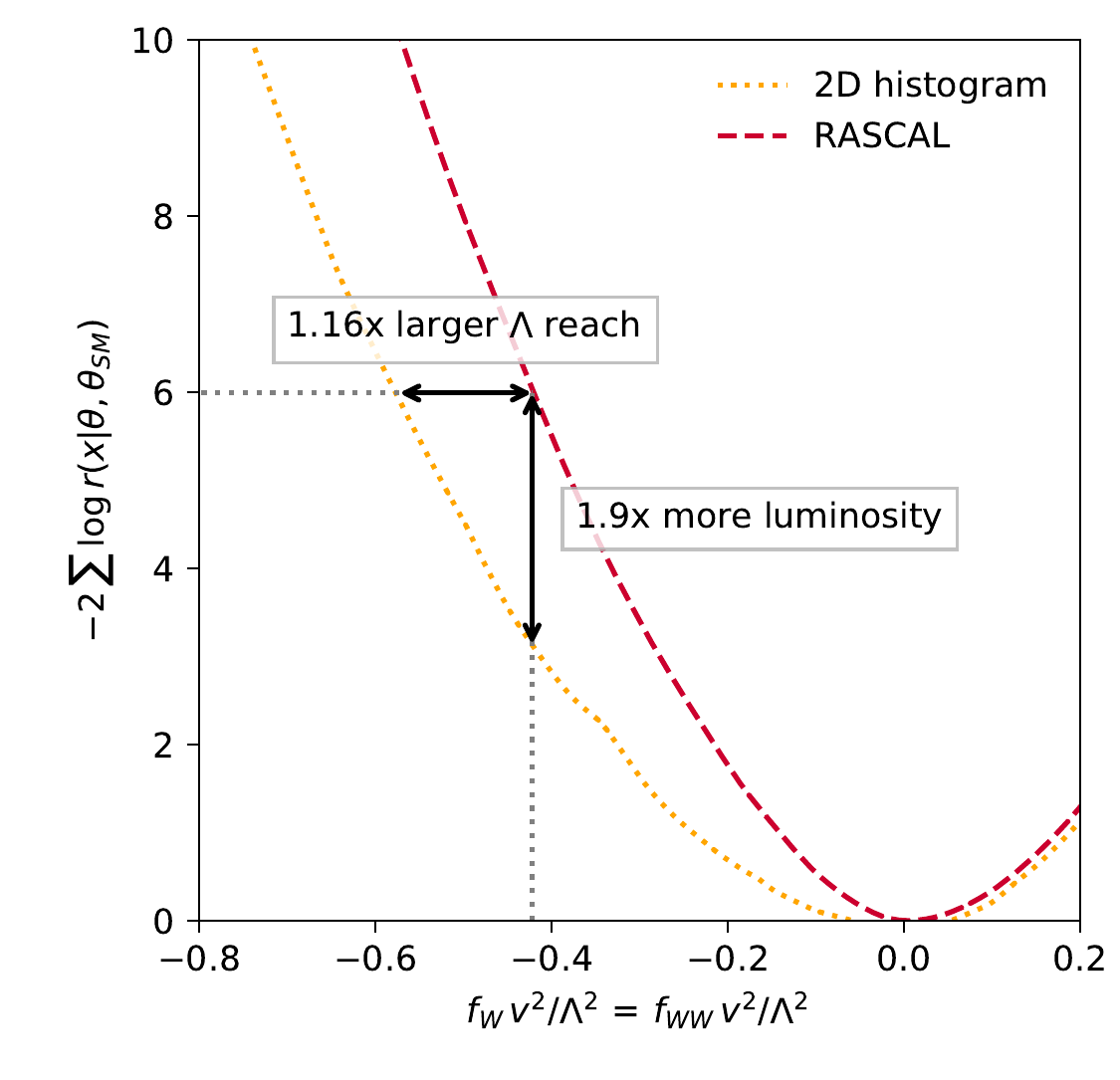}%
  \includegraphics[width=0.49 \textwidth]{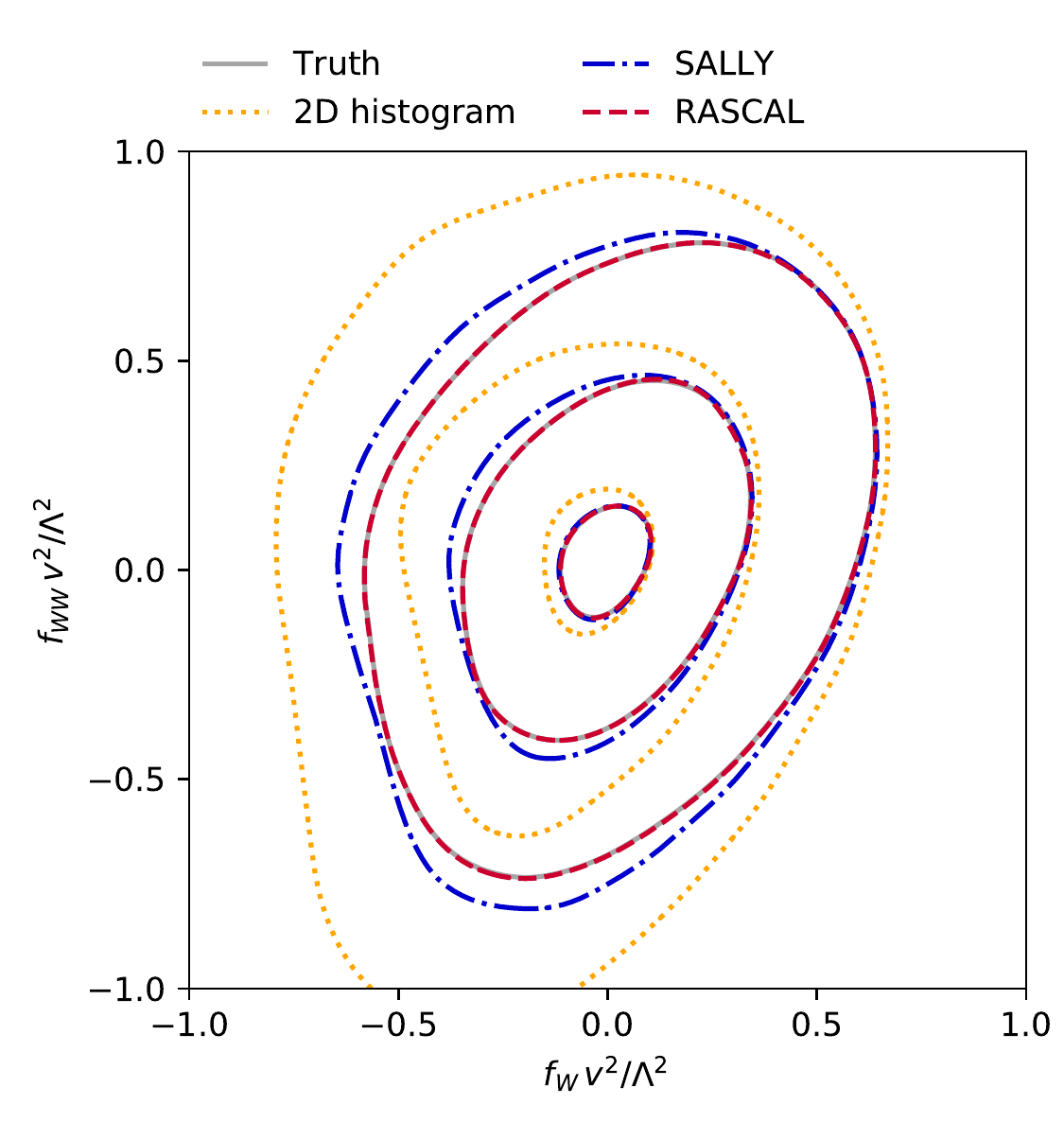}%
  \caption{Left: Estimated expected likelihood ratio based on a traditional doubly differential histogram analysis (orange dotted) and the new \rascal technique (red dashed). We show a line in parameter space with particularly large difference between the methods. The grey dotted line marks the expected exclusion limit at $95 \%$ CL according to asymptotics. The vertical arrow shows how much more data the histogram approach requires to constrain the same parameter point with the same significance. The horizontal arrow demonstrates the increased physics reach of the machine-learning-based method. Right: Expected exclusion contours at $68\%$~CL (innermost lines), $95\%$~CL, and $99.7\%$~CL (outermost lines) based on the Neyman construction. In both panels, we assume 36 observed events and the SM to be true.}
  \label{fig:results}
\end{figure}

In the left panel of Fig.~\ref{fig:results} we show the approximate likelihood ratio estimated with the \rascal method for one particular slice through parameter space. We also show the likelihood ratio based on a traditional histogram-based analysis of two particularly powerful kinematic variables, the transverse momentum of the hardest jet and the azimuthal angle between the two jets~\cite{Brehmer:2016nyr, Brehmer:2017lrt}. The new method clearly enables stronger exclusion limits, equivalent to a $16\%$ larger reach in the new physics scale or $90\%$ more collected data in this particular parameter region.

The right panel of Fig.~\ref{fig:results} shows expected constraints on the two operators after 36 observed events with the \rascal and \sally methods based on the Neyman construction. The results for \sallino are very similar to those for \sally. The \rascal limits are virtually indistinguishable from the true likelihood contours. \sally and \sallino lead to nearly optimal bounds close to the Standard Model, slightly weaker constraints at the $95\%$~CL level show the breakdown of the local model approximation. All new techniques let us impose significantly tighter bounds on the parameters than the doubly differential histogram analysis.

\section{Conclusions}
\label{sec:conclusions}

We have developed new analysis techniques to constrain effective field theories in LHC experiments. Exploiting the particular structure of particle physics processes, we extract additional information from Monte-Carlo simulations. This augmented data can be used to train neural networks that estimate arbitrary likelihood ratios for use in limit setting procedures.

We have introduced the \rascal technique, which leverages this extended information to define likelihood ratio estimators of particularly high fidelity. In an example analysis of weak-boson-fusion Higgs production, this technique lets us put significantly stronger constraints on two dimension-six operators, leading to expected exclusion limits that are virtually indistinguishable from the theoretical optimum.

In the neighborhood of the Standard Model, any observation can be condensed into a low-dimensional vector, the score, without loss of sensitivity. This motivates a second approach, which we call \sally. Simpler to implement, it scales very well to high-dimensional parameter spaces. We have demonstrated that it performs very well close to the Standard Model, and leads to only slightly weaker constraints further away.

Both approaches scale well to large-scale LHC analyses with many observables and high-dimensional parameter spaces. Though the new methods are particularly well-suited to the SMEFT, they can be applied more generally. They do not require any approximations of the hard process, parton shower, or detector effects, and the likelihood ratio can be evaluated in microseconds. Given their performance, scalability, and practicality, these techniques have the potential to substantially improve the LHC legacy measurements.

\subsection*{Acknowledgments}

We would like to thank Cyril Becot and Lukas Heinrich, who contributed to this project at an early stage. We are grateful to Felix Kling, Tilman Plehn, and Peter Schichtel for providing the \toolfont{MadMax} code and helping us use it. KC wants to thank CP3 at UC Louvain for their hospitality. Finally, we would like to thank At{\i}l{\i}m G\"{u}ne\c{s} Baydin, Lydia Brenner, Joan Bruna, Kyunghyun Cho, Michael Gill, Ian Goodfellow, Daniela Huppenkothen, Hugo Larochelle, Yann LeCun, Fabio Maltoni, Jean-Michel Marin, Iain Murray, George Papamakarios, Duccio Pappadopulo, Dennis Prangle, Rajesh Ranganath, Dustin Tran, Rost Verkerke, Wouter Verkerke, Max Welling, and Richard Wilkinson for interesting discussions.

JB, KC, and GL are grateful for the support of the Moore-Sloan data science environment at NYU. KC and GL were supported through the NSF grants ACI-1450310 and PHY-1505463. JP was partially supported by the Scientific and Technological Center of Valpara\'{i}so (CCTVal) under Fondecyt grant BASAL FB0821. This work was supported in part through the NYU IT High Performance Computing resources, services, and staff expertise.


\bibliography{references}

\end{document}